\documentstyle[aps,multicol,epsfig]{revtex}
\begin{document}
%
\author{D.H.Cobden\cite{Berkeley} \and B.A.Muzykantskii}
\address{Cavendish Laboratory, Madingley Road, Cambridge CB3 0HE, UK}
\title{Finite-temperature Fermi-edge singularity in tunneling
studied using random telegraph signals}
\date{April 19, 1995}
\draft
\maketitle
\widetext
\begin{abstract}
We show that random telegraph signals in metal-oxide-silicon
transistors at millikelvin temperatures provide a powerful means of
investigating tunneling between a two-dimensional electron gas and a
single defect state.  The tunneling rate shows a peak when the defect
level lines up with the Fermi energy, in excellent agreement with
theory of the Fermi-edge singularity at finite temperature.  This
theory also indicates that defect levels are the origin of the
dissipative two-state systems observed previously in similar devices.
\end{abstract}
%
\begin{multicols}{2}
In small electrical devices, noise signals are often seen which
reflect the transitions of a single atom or electron between two or
more metastable states.  These `random telegraph signals' (RTSs)
\cite{Ralls1}, where the conductance jumps randomly in time between
certain discrete values, are being increasingly exploited as a means
of investigating a diverse range of tunneling phenomena, such as
dissipative tunneling of two-state systems \cite{Golding},
electromigration \cite{Ralls2}, hopping conduction \cite{Cobdenhop},
and tunneling between quantum Hall edge channels \cite{van der Vaart}.
In the present work we use them for the first time to study the
dynamics of electrons tunneling between a defect state and a
two-dimensional electron gas (2DEG) at millikelvin temperatures.  We
find that a noninteracting electron picture cannot explain the
behaviour of the system.  This is not very surprising, because the
interaction phenomena of Coulomb blockade \cite{Beenakker} and the
Kondo effect \cite{Hershfeld} are known to strongly influence the
tunnel conductance through a single defect level \cite{Dellow,Geim} or
a quantum dot \cite{Reed}.  However, what is surprising is that the
dominant interaction effect in the present system is another, to which
attention has only recently been drawn \cite{Matveev}.  It is the
interaction of the 2D gas electrons with the defect potential, which
produces a peak in the tunneling rates near the Fermi level at low
temperature $T$.  At $T=0$ this peak becomes a Fermi-edge singularity,
whose origin is the same as that of the X-ray absorption edge
singularity in metals \cite{Mahan}.

This singularity has already been invoked to explain sharp peaks
observed in the differential conductance of small tunnel barriers
\cite{Geim}, but the interpretation in that situation was hampered by
the lack of equilibrium due to the large voltage bias, and by the
simultaneous presence of other anomalous structure in the device
characteristics.  In our RTS experiments we can measure the tunneling
rates for an isolated defect directly and in thermal equilibrium.  We
find remarkable agreement with the theory for the finite temperature
generalization of the Fermi-edge singularity, which has not been
tested experimentally before.  Using the same theory we are then able
to attribute the weakly coupled dissipative two-state systems observed
previously \cite{CobdenTSS} in similar devices to an electron
tunneling between a defect level and a bound state of the defect
potential.  We also find that the effects of a magnetic field $B$ on
an RTS are fully consistent with the electron-trapping defect
scenario.

The measurements were made on two-terminal Si
metal-oxide-semiconductor field-effect transistors (MOSFETs) with
highly doped contact regions, oxide thickness $d_{ox} = 240$ \AA, and
channel dimensions 0.6 or 0.8 $\mu$m.  Care was taken not to stress
the devices electrically, and the threshold gate voltage was about 2.0
V in a dilution refrigerator at 100 mK.  The resistance was sampled at
up to 5 kHz using a standard constant-current lock-in technique with a
Brookdeal 5004 ultra-low noise voltage preamplifier.  The
low-temperature peak mobility was around 0.2 m$^2$V$^{-1}$s$^{-1}$,
corresponding to a transport scattering length of $ \sim 300$ \AA.
This rather high disorder leads to quantum interference effects (see
later) which fortuitously make the RTSs big enough to allow the use of
signal levels not significantly larger than $kT/e$ (to avoid resistive
heating) down to $100$~mK.  The mean time spent in each resistance
state of an RTS under fixed conditions was found by averaging over
several hundred transitions, giving a standard error of a few percent.

Previous work at $T \ge 4.2$ K has shown that most RTSs in n-channel
MOSFETs result from defect levels situated in the oxide close to the
Si/SiO$_2$ interface \cite{Ralls1}.  Fig.~1(a) shows a schematic band
diagram illustrating the situation at low $T$, together with a section
of a typical RTS.  A positive voltage $V_g$ is applied to the gate,
sufficient to create a degenerate 2DEG, represented by the shaded
area at the interface.  $E_d$ is a defect level (indicated as being in
the oxide, though its location is not important here), $E_F$ is the
Fermi level, and $E_0$ is the bottom of the lowest 2D subband.  We
denote the conductance of the device with the defect level empty
(state 1) or occupied (state 2) by $G_1$ or $G_2$ respectively, and
the reciprocals of the mean times spent in the two states by rates
$\gamma_1$ and $\gamma_2$, as indicated in the figure.  The ratio of
these rates satisfies the detailed balance condition,
\begin{equation}
\frac{\gamma_1}{\gamma_2}   =   \exp[-(E_d-E_F)/kT] .
\label{dbalance}
\end{equation}
The lower graphs in Fig.~1(b) show $\ln(\gamma_1/\gamma_2)$ plotted
against $V_g$ at two temperatures.  The variation is almost linear,
implying that
\begin{equation}
E_d-E_F = -\eta e(V_g-V_{g0}),
\label{E vs Vg}
\end{equation}
where $\eta$ (the `sensitivity') and $V_{g0}$ (the `balance' gate
voltage) are constants.  The straight solid lines in the figure
correspond to $\eta = 0.019$ and $V_{g0} = 6.662$~V.  The linearity
results from the energy-independent density of states in the inversion
layer and the linear sensitivities of $E_d$ and $E_0$ to the oxide
electric field over the relevant range of $V_g$.  Differences in
$\eta$ between individual RTSs can be attributed to different defect
locations.  The deduced values of $E_d-E_F$ for RTS\#1 are plotted
along the top axes in Fig.~1(b).

In a noninteracting picture the individual rates, deduced from the
golden rule, are given by
\begin{equation}
\begin{array}{rcl} \gamma_1 &=& \frac{2\pi}{\hbar} D \Delta^2 f(E_d)
\\[1ex] 
\gamma_2 &=& \frac{2\pi}{\hbar} D \Delta^2 \left[1-f(E_d) \right],
\end{array}
\label{golden rule}
\end{equation}
where $f(E)$ is the Fermi function, $D$ is the electron density of
states, and $\Delta$ is the tunneling matrix element.  The lines on
the upper graphs in Fig.~1(b) are plots of Eqs.~(\ref{golden rule}),
where $(2\pi/\hbar) D \Delta^2 = 280$ s$^{-1}$ is the only fitting
parameter.  At $T = 1.2$ K the measured values of $\gamma_1$ and
$\gamma_2$ follow them rather well; hence the picture of tunneling
without interactions seems to suffice.  However, when $T$ is reduced
to $0.5$ K a distinct peak appears in the rates in the region of $E_d
= E_F$.  To account for this peak, which occurs to some degree for
every RTS, we are forced to go beyond the noninteracting picture.
Matveev and Larkin \cite{Matveev} have recently pointed out that one
should take into account the consequences of the change in the defect
potential seen by electrons in the 2DEG when an electron tunnels.
Most of the relevant theory was developed in the context of the X-ray
absorption edge \cite{Mahan}, and at $T=0$ it predicts a power-law
singularity in the transition rate,
\begin{equation}
       \gamma_1 \sim \theta(E_F - E_d) \left( E_F -E_d \right)^{\alpha-1},
\label{formula0}
\end{equation}
where $\theta (E)$ is the unit step function.  The singularity arises
because the tunneling electron can easily lose a small amount of
energy to low-energy electron-hole pairs which are created by the
sudden change in the defect potential.  The finite temperature
generalization of Eq.~(\ref{formula0}) is \cite{Anderson}
\begin{eqnarray}
   \gamma_{1,2} & = & CT^{\alpha-1}
\exp\left( \pm \frac{E_F-E_d}{2kT} \right)
\nonumber \times \\
&& \qquad \frac{|\Gamma[\alpha/2+i(E_d-E_F)/(2\pi kT)]|^2}{\Gamma(\alpha)},
\label{formula}
\end{eqnarray}
where $C$ is a constant and $\alpha$ is equal to the zero-temperature
exponent.  Fig.~2 shows the transition rates for RTS\#2, found in the
same device as RTS\#1 at lower gate voltage ($V_{g0} = 3.0863$ V and
$\eta = 0.018$).  The solid lines are plots of Eq.~(\ref{formula})
using $\alpha = 0.21 \pm 0.01$ and $C = 10.2 \pm 0.2$ s$^{-1}$ (with
$T$ in degrees Kelvin).  The very good fits at both T = 145 mK and 360
mK are convincing evidence that these measurements directly probe the
Fermi-edge singularity at finite temperature.

In the theory, $\alpha$ depends only on the defect potential. For
simplicity, let us assume the potential before capture, $V$, is
attractive and radially symmetric, while the potential after capture
is zero.  The one-electron eigenstates can be classified by the
perpendicular component of the angular momentum $m=0,\pm1,\pm2\ldots$,
together with spin and valley indices.  In the 2-DEG at $B=0$ there
are two equivalent conduction band valleys, and for convenience we
combine spin and valley into a single index $s={1\ldots4}$.  If the
phase shift for channel $(m,s)$ at $E_F$ in the presence of $V$ is
$\delta_{m,s}$, then $\alpha$ is given by
\cite{Combescot}
\begin{equation}
\alpha = \left(-1+\frac{\delta_{0,1}}{\pi}\right)^2 +
         \sum_{(m,s)\ne(0,1)} \left( \frac{\delta_{m,s}}{\pi}\right)^2.
\label{alpha}
\end{equation}
Note that if $V=0$, $\delta_{m,s}=0$ for all $m$ and $s$, $\alpha=1$, 
and with the correct value of $C$ Eq.~(\ref{formula}) reduces 
to the noninteracting result, Eq.~(\ref{golden rule}).  
Charge neutrality forces the phase shifts to obey the Friedel sum rule
\cite{Friedel}, $\sum_{m,s} \delta_{m,s} = \pi$, and we can use this
together with Eq.~(\ref{alpha}) to obtain a lower limit on $\alpha$.
This limit is reached for pure $s$-wave scattering, when
$\delta_{m,s}=0$ for all $m\neq 0$.  In 2-D an attractive potential
always has a bound state.  When screening is strong, ie, for large
$E_F$, the bound state may be occupied by four electrons
simultaneously \cite{Vinter} and one finds $\alpha \ge 3/4$.  For
weaker screening (small $E_F$) the bound state is occupied by only one
electron and one finds $\alpha \ge 0$.  The limit $\alpha = 0$ is
reached when the electron tunnels directly to the defect level from
the bound state while all extended states remain completely
unaffected, ie, all the screening is done by the single electron in
the bound state.  Then $\delta_{m,s}$ is zero for all channels except
the one in which the bound state was destroyed, whose phase shift is
$\delta_{0,1}=\pi$.

The value $\alpha=0.21$ obtained for RTS\#2 is only consistent with a
singly occupied bound state, with $\pi/2 < \delta_{0,1} < \pi$ and
fairly small phase shifts in other channels.  On the other hand, the
values $\alpha \sim 0.7$ and $0.9$ for RTSs \#1 and \#3 respectively
allow the possibility of multiple occupancy of the bound state.  At
this point it is interesting to reconsider the results of some earlier 
RTS experiments in MOSFETs \cite{CobdenTSS,Schulz}.  In
Ref. \cite{CobdenTSS} the data were fitted using an expression
identical in form to Eq.~(\ref{formula}) but derived from the theory
of two-state systems (TSSs).  This remarkable identity is in fact no
coincidence, because an electron tunneling between a defect level and
a single bound state in the defect potential constitutes a TSS.  If
one could vary $\alpha$ from 0 to 1 one could in principle transform
the system continuously from a noninteracting TSS into a
noninteracting defect level.  While the defects in the present work
are in the intermediate regime, those in Ref. \cite{CobdenTSS} were
found to have $\alpha \ll 1$, corresponding to TSSs very weakly
coupled to the environment (the 2DEG).  In this limit
Eq.~(\ref{formula}) takes the dramatically different form of a narrow
Lorentzian centered at $E_d=E_F$.  We suggest that for such defects,
which are seen only in electrically stressed devices, the 2DEG is
locally depleted out due to potential fluctuations at the interface.
The extended states therefore remain unaffected on electron capture
>from the bound state because they are distant from the
defect/bound-state system.  For these defects we can also offer a
resolution of a paradox that would arise if the defect were located in
a metallic region, namely, that a small value of $\alpha$ implies very
weak scattering of electrons at $E_F$, while a large RTS amplitude
appears to require strong scattering.  If the defect actually lies in
a depleted region then electrons at $E_F$ may be able to tunnel across
this region via the bound state when the defect is ionized.  Hence the
conductance decreases by as much as $e^2/h$ when the defect captures
an electron, because the bound state is destroyed, even though all
extended-state wavefunctions are unaltered.

Finally, we examine the effects of a magnetic field on another RTS,
RTS\#3 (seen in a different device from RTS\#1), which lend further
support to the basic picture of electrons tunneling between a defect
and the 2DEG.  For RTS\#3, $\alpha \sim 0.9$, so the deviations from
the noninteracting result, Eq.~(\ref{golden rule}), are moderate.  In
Fig.~3, trace (i) is the variation of $G_1$ up to $12$ T.  The edge of
a quantum Hall plateau is visible at the highest field, while only
universal conductance fluctuations \cite{Lee} are seen at lower $B$.
Trace (ii) shows the corresponding variation of $V_{g0}$, which above
about $4$ T undergoes oscillations commensurate with the vertical
dotted lines marking points at which $n$ Landau levels are full.  This
confirms our expectation that the tunneling process should be
sensitive to the modulations of the density of states in the 2DEG,
although the nature of this sensitivity is complex and will be the
subject of future work.  Inset trace (iii) is an expanded section of a
magnetic field sweep made whilst sampling the conductance at $1$ kHz,
illustrating how $G_1$ and $G_2$ undergo distinct reproducible
fluctuations.  Trace (iv) shows $\Delta G_1$, which is the deviation
of $G_1$ from the smooth classical magnetoresistance background
present in trace (i).  Trace (v) shows the RTS amplitude $\delta
G_{21} \equiv G_2-G_1$ over the same field range.  As can be seen, the
fluctuations in $\delta G_{21}$ and $\Delta G_1$ are qualitatively
similar and exhibit virtually the same correlation length $B_c$
\cite{Lee}.  In the range $0 < B < 5$~T we find $B_c = 0.04 \pm
0.01$~T, giving an estimate of $A_\phi = (B_c e/h)^{-1} = 0.1 \; \mu
$m$^2$ for the typical phase-coherent area, compared with the device
area $A \sim 0.5 \; \mu $m$^2$.  The rms amplitudes of the
fluctuations in the two quantities are $\sigma_1 = 10 \pm 1 \; \mu $S
and $\sigma_{21} = 3.6 \pm 0.4 \; \mu $S.  Hence $\sigma_{21}/\sigma_1
\sim 0.36 \pm 0.06$.  Allowing for the uncertainties in $A$ and
$A_\phi$, this is consistent with the prediction
$\sigma_{21}/\sigma_1\sim (A_\phi/A)^{1/2} \sim 0.45 $ for the removal
of a single strong scatterer from a disordered 2D system at finite
temperature \cite{Feng,Golding}.  The dashed line superimposed on
trace (v) was obtained by smoothing $\delta G_{21}$ over a range of
$2$ T.  Notice that it is always above zero, ie, on average $G_2 >
G_1$.  This helps to justify our earlier simplifying assumption that
on electron capture a scattering center is removed from the 2DEG, ie,
that the defect goes from positive to neutral.

In conclusion, using random telegraph signals we have shown that
electronic tunneling between a defect level and a 2DEG exhibits the
finite-temperature counterpart of the Fermi-edge singularity, with
exponent $\alpha$ ranging from nearly zero to nearly one.  Defects 
which have very small values of $\alpha$ behave as two-state systems.

We thank C. Barnes, P. McEuen, M. Pepper and M. J. Uren for helpful
discussions, and Y. Oowaki of Toshiba for supplying the devices.

\end{multicols}

\begin{figure}
\raisebox{3.7cm}{%
\makebox[0.48\textwidth]{\psfig{figure=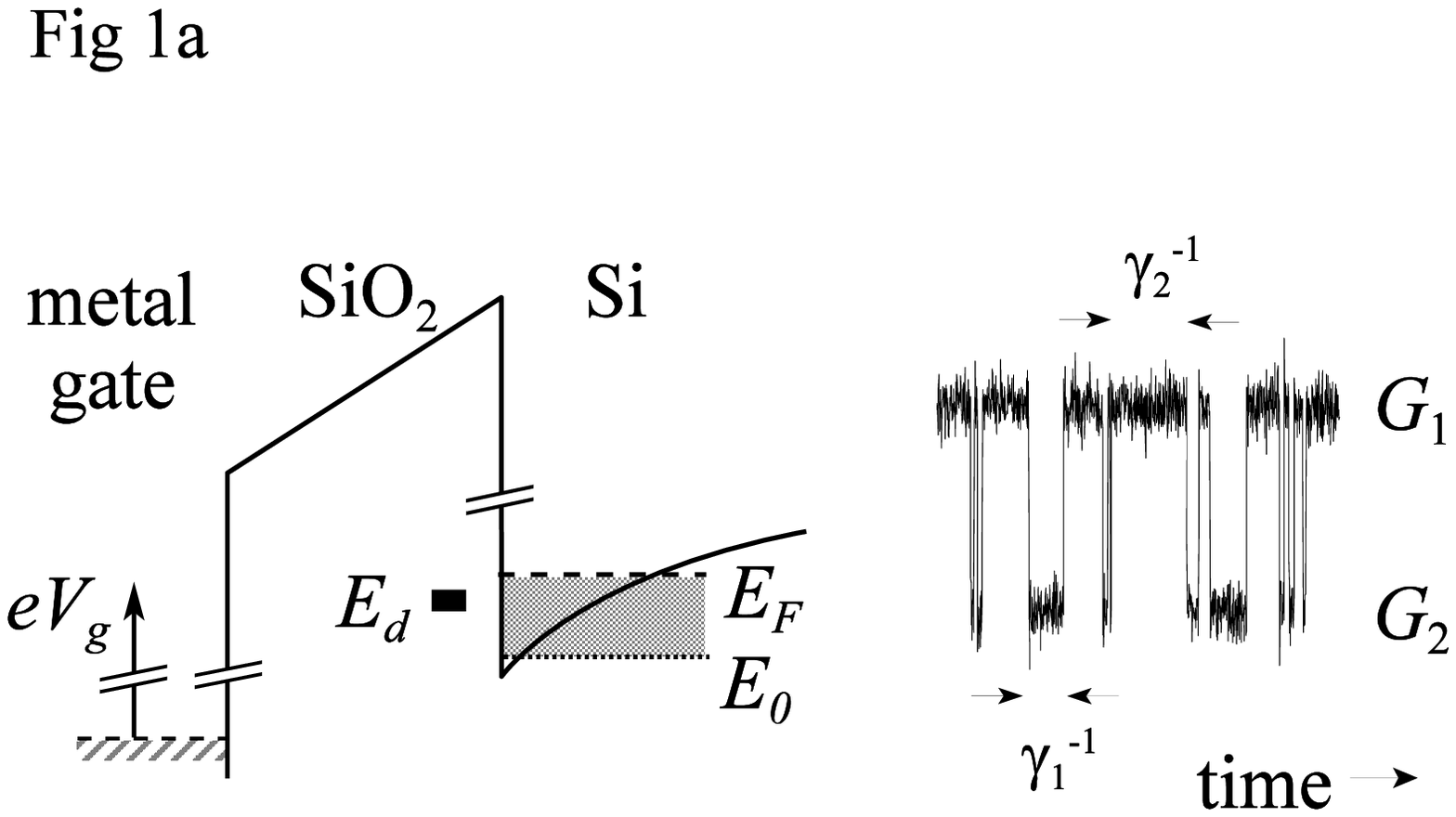,width=0.48\textwidth}}}
\makebox[0.48\textwidth]{\psfig{figure=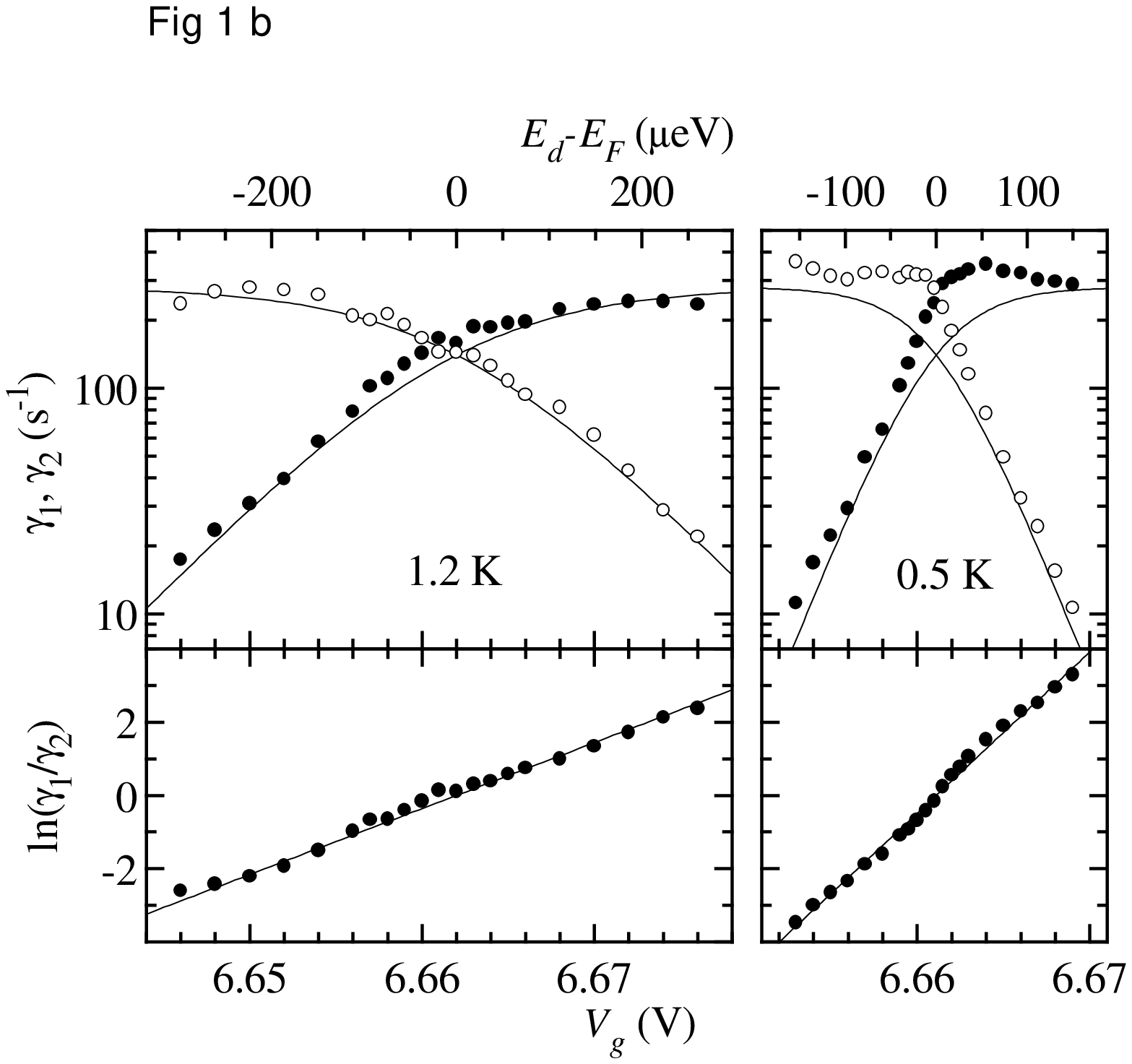,width=0.48\textwidth}}
\caption[]{(a) Schematic band diagram of a MOSFET at low 
temperature, and a section of a typical RTS seen in such a device.
(b) Gate-voltage dependence of the capture rate $\gamma_1$ (filled
circles) and emission rate $\gamma_2$ (empty circles) for RTS\#1 at
$1.2$ K and 0.5 K, with $B=0.28$ T.  The device
conductance was $2$ mS.  The solid lines are fits to
Eqs.~(\ref{dbalance}) - (\ref{golden rule}), which neglect
interactions.}
\label{fig1}
\end{figure}

\begin{figure}
\centerline{\psfig{figure=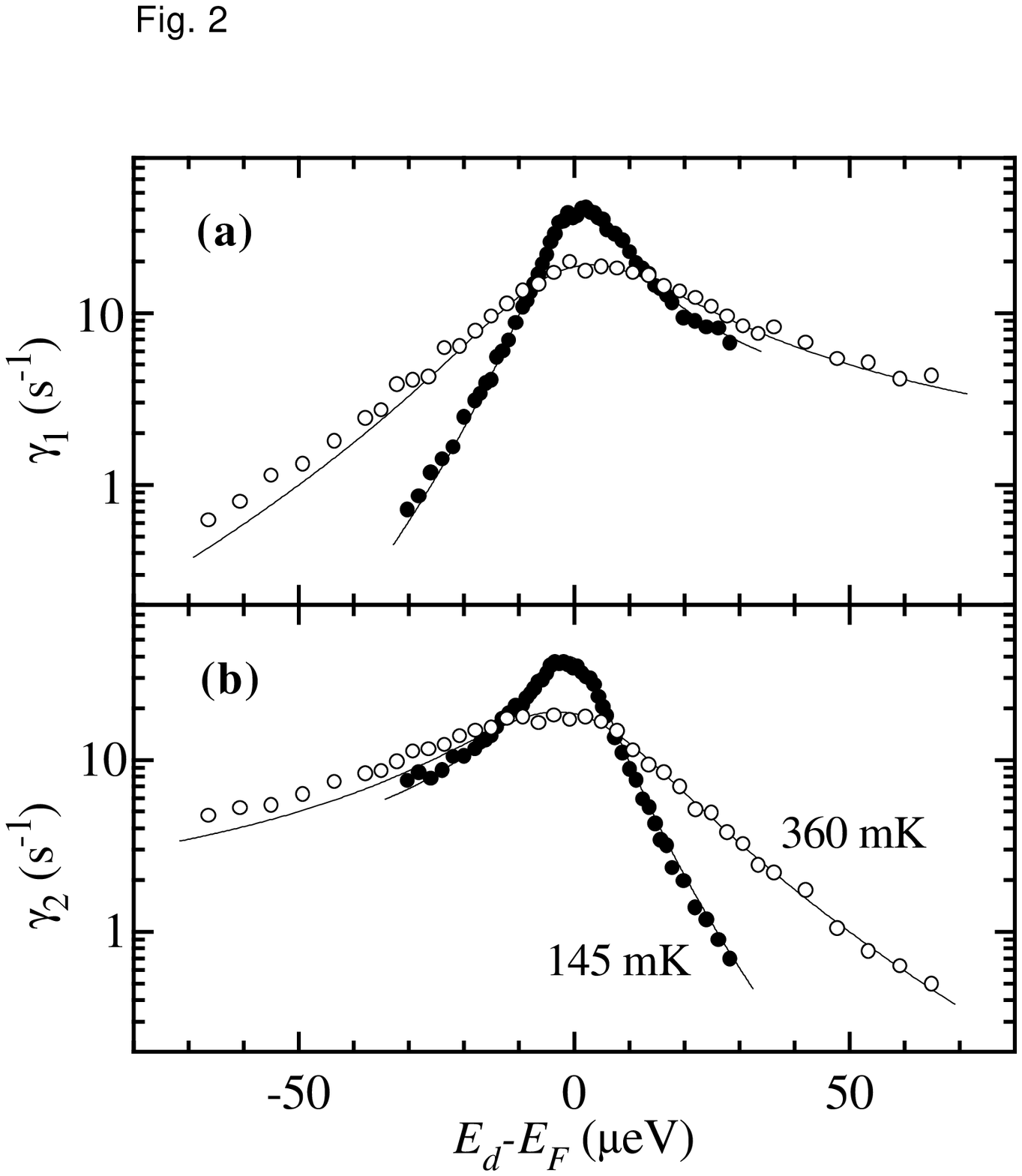,height=0.37\textheight,clip=}}
\caption[]{Energy dependence of (a) capture and (b) emission rates for RTS\#2 
at $145$ mK (filled circles) and $360$ mK (empty circles) at $B =
0.06$ T.  The solid lines are fits to Eq.~(\ref{formula}) (see text).}
\label{fig2}
\end{figure}

\begin{figure}
\centerline{\psfig{figure=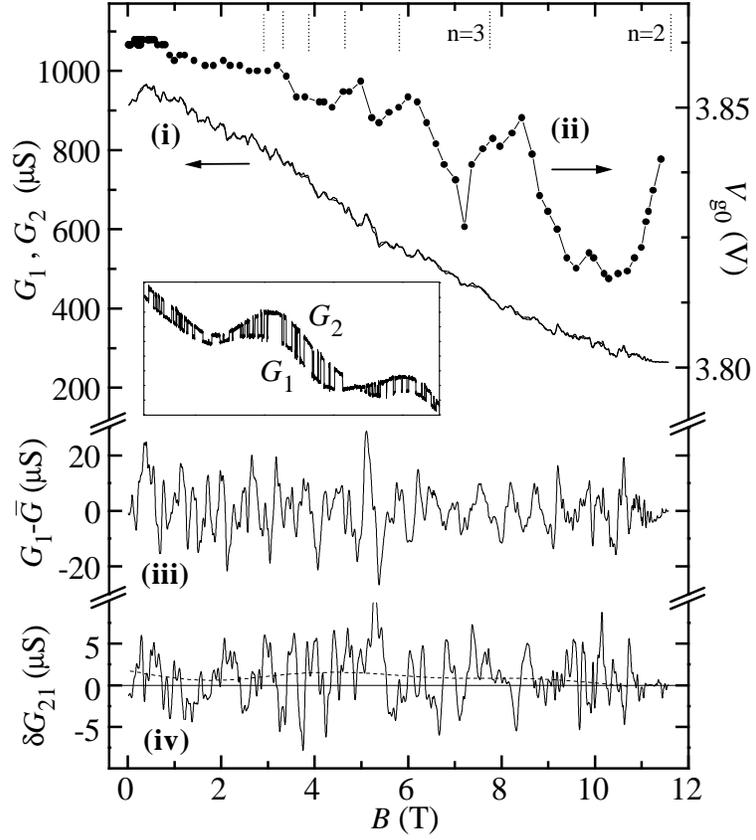,height=0.5\textheight}}
\caption[]{Effects of magnetic field on RTS\#3 at 100 mK, for 
which $\eta= 0.007$ at $B=0$. Trace (i) (left axis), conductance
$G_1$ in state $1$ of the RTS.  Trace (ii) (right axis), balance
gate voltage Vg0, showing oscillations commensurate with Landau level
index $n$.  Trace (iii) (inset), sweep of magnetic field over a small
range at high bandwidth, where both levels of the RTS are visible.
Traces (iv) and (v), variation of $\Delta G_1$ and $\delta G_{12}$
over the full field range.}
\label{fig3}
\end{figure}

\end{document}